\documentclass[aps,pre,twocolumn,twoside,floatfix,a4paper,showpacs]{revtex4}

\usepackage{bbm}
\usepackage{amsmath, amssymb}
\usepackage{color}
\usepackage{graphicx,epsfig}
\usepackage{dcolumn}
\usepackage{amsmath}
\usepackage{pstricks}
\usepackage{psfrag}
\usepackage{subfigure}

\newcommand{\ew}[1]{\langle #1 \rangle}
\newcommand{\tr}[1]{\operatorname{tr}\!\left\{ #1 \right\}}
\def\<{\langle}
\def\>{\rangle}
\newcommand{\eq}[1]{Eq.~\eqref{#1}}
\DeclareMathOperator{\arcoth}{arcoth}

\begin{document}

\title{Hamiltonian of mean force for  damped quantum systems}

\author{Stefanie Hilt, Benedikt Thomas, and Eric Lutz}
\affiliation{Department of Physics, University of Augsburg, 86135 Augsburg, Germany}

\date{\today}

\begin{abstract}
We consider a quantum system linearly coupled to a reservoir of harmonic oscillators. For  finite coupling strengths, the stationary distribution of the damped system is not of the Gibbs form, in contrast to standard thermodynamics. With the help of the quantum Hamiltonian of mean force, we quantify this deviation exactly for a harmonic oscillator and provide approximations in the limit of high and low temperatures, and weak and strong couplings. Moreover, in the semiclassical regime, we use the quantum Smoluchowski equation to obtain results valid for any potential. We, finally, give a physical interpretation of the deviation in terms of the  initial system-reservoir coupling.
\end{abstract}

\pacs{03.67.-a, 05.30.-d}

\maketitle

\section{Introduction}\label{sec:1}
Thermodynamics successfully describes equilibrium properties of macroscopic objects coupled to a heat bath.   In the standard theory, the system-bath interaction is assumed to be small and is therefore  neglected. This is a well-justified assumption when the coupling energy is small compared to the thermal energy of the system \cite{kub68}.  However, at low temperatures, quantum effects come into play and the interaction energy cannot be discarded anymore \cite{han05}. This is particularly relevant for solid-state nanodevices, such as nanomechanical oscillators, which may be strongly coupled to their environment \cite{cle02,rou01,reg08,kno03}. In the limit of vanishingly small coupling, the damped quantum system asymptotically relaxes to the correct thermal Gibbs state \cite{ben81}. However, for any finite interaction strength, the stationary state of the quantum system deviates from  a Gibbs state, in contrast to the prediction of thermodynamics. It has, for example,  been shown that, at zero temperature, the coupled oscillator is in an excited, mixed state and not in its pure ground state \cite{lin84,li95,for06}. In addition, it is worth noticing that the deviation from a Gibbs state leads to the interesting possibility of entangling two noninteracting coherent states via the indirect coupling to a common reservoir \cite{paz08}.

A useful quantity to describe classical systems that are strongly coupled to their environment is the so-called potential of mean force \cite{tuc10}. It was first introduced by Kirkwood in his study of the average structure of liquids \cite{kir35}. It has since proved to be an invaluable tool in chemistry \cite{hil56}, where  it is commonly employed in the investigation of implicit solvent models \cite{roux99} and of protein-ligand binding \cite{rod08}, to name a few examples. A quantum extension in the form of the  Hamiltonian of mean force (HMF), see \eq{Eq:2} below, has recently been proposed in Ref.~\cite{cam09}. Potential and Hamiltonian of mean force are instrumental in characterizing the deviations from standard thermodynamics that occur in the strong coupling regime. However, they are difficult to evaluate explicitly in the general case.
Here, we use the quantum HMF to quantify the deviation from the  thermal Gibbs state for a damped quantum harmonic oscillator, for which we obtain exact analytical results. We further derive approximate expressions for arbitrary systems in the strong damping  regime. We additionally relate  the HMF to the thermodynamic change of energy and entropy associated with the initial coupling between system and bath.

The present article is organized as follows. In Section \ref{sec:2}, we introduce the concept of the quantum HMF and show how it can be used to characterize the deviation from a thermal Gibbs state. In Section \ref{sec:3}, we determine the exact expression for this deviation for the case of the analytically solvable damped quantum harmonic oscillator. In Section \ref{sec:4}, we provide useful approximations in the limit of  high  and low temperatures,   weak and strong coupling. Furthermore in the semiclassical limit, we employ the quantum Smoluchowski equation to evaluate the deviation from the Gibbs state for any  damped quantum system.   In Section \ref{sec:5}, we finally give a physical interpretation of the HMF by establishing a relationship with the initial system-bath coupling. 
\section{Hamiltonian of mean force}\label{sec:2}
We consider the total Hamiltonian,
\begin{align}
	H = H_S+H_B+H_{SB}\label{Eq:1}\ ,
\end{align}
which describes a quantum system $S$ coupled to a heat bath $B$ via the interaction $SB$. The equilibrium state of the total system is given by the Gibbs state,
\begin{align}
	\rho=\frac{\exp{(-\beta H)}}{Z}\ ,
\end{align}
where $Z=\tr{\exp(-\beta H)}$ is the total partition function and $\beta=1/(k T)$ the inverse temperature of the bath.
When  the interaction energy is negligible, $H \simeq H_S+H_B$, the reduced density operator of the system, $\rho_S=\mbox{tr}_B\{\rho\}$,
is also of the Gibbs form,
\begin{align}
	\rho_S= \frac{\exp{(-\beta H_S)}}{Z_S}\ ,\label{Eq:d}
\end{align}
with $Z_S=\mbox{tr}_S\{\exp(-\beta H_S)\}$.
However, for finite coupling, the interaction term cannot be neglected, and the reduced density operator  is no longer  of the Gibbs form. It can generically be written as,
\begin{align}
	\rho_S= \frac{\exp{(-\beta H^*_S)}}{Z^*}\label{Eq:2a} \ ,
\end{align}
where  the quantum HMF is defined as the effective Hamiltonian \cite{cam09},
\begin{align}
	H^*_S= -\frac{1}{\beta} \ln \frac{\mbox{tr}_B\{\exp(-\beta H)\}}{ \mbox{tr}_B\{ \exp(-\beta H_B)\}}\ .\label{Eq:2}
\end{align}
The corresponding partition function is given by $Z^*=\mbox{tr}_S\{\exp(-\beta H^*_S)\}$. The HMF reduces to the Hamiltonian of the system, $H^*_S\simeq H_S$, in the limit of vanishing  coupling. It is thus convenient to introduce the difference,
\begin{align}
	\Delta H_S =H^*_S - H_S\label{Eq:3}\ .
\end{align}
The latter  quantifies the deviation from a thermal Gibbs state. In the following, we evaluate the deviation $\Delta H_S$ explicitly for the damped quantum harmonic oscillator.

\section{Microscopic system-reservoir model}\label{sec:3}
The standard model for a damped quantum harmonic oscillator is given by Hamiltonian (\ref{Eq:1}) \cite{wei99}, where 
\begin{align}
	H_S=\frac{p^2}{2M}+ \frac M 2 \omega^2 q^2
\end{align}
is the Hamiltonian of the harmonic oscillator with mass $M$ and frequency $\omega$,
\begin{align}
	H_B=\sum_{j=1}^N\left[\frac{p_j^2}{2 m_j}	+\frac{m_j\omega_j^2}{2}x_j^2\right]
\end{align}
describes the bath of $N$ harmonic oscillators, and
\begin{align}
	H_{SB} =\sum_{j=1}^N \left[-C_j q x_j+\frac{C_j^2}{2m_j\omega_j^2}q^2\right]\label{Eq:4}\ .
\end{align}
is the  coupling term linear in the position of the system. The frequencies of the reservoir modes are taken to be equidistant, that is $\omega_j = j \Delta$. The coupling constants $C_j$ are chosen to obey a Drude-Ullersma spectrum \cite{wei99},
\begin{align}
	 C_j=\sqrt{\frac{2 \gamma m_j M \omega_j^2 \Delta}{\pi}\frac{\omega_D^2}{\omega_D^2+\omega_j^2}}\,,\label{Eq:5}
\end{align}
with damping coefficient $\gamma$ and  cutoff frequency $\omega_D$. In the continuous limit $N\rightarrow\infty$ ($\Delta\rightarrow 0$), the bath is  characterized by the Ohmic spectral density function \cite{wei99},
\begin{align}
	J(\nu)=\frac{\pi}{2}\sum_j \frac{C_j^2}{m_j \omega_j}\,\delta(\nu-\omega_j)= \frac{\gamma \nu M \omega_D^2}{\nu^2+\omega_D^2}\ .\label{Eq:6}
\end{align} 
Due to the linearity of the model, the dynamics of the reduced quantum system is exactly solvable. Its stationary state   is conveniently described in phase space using the  Wigner function \cite{sch01},
\begin{align}
	W(q,p,t)=\int_{-\infty}^\infty \frac{d \xi}{2\pi\hbar}\,\exp{\left[-\frac{\mbox{i}}{\hbar}p\xi\right]}\rho_S\left(q+\frac\xi 2,q-\frac\xi 2,t\right)\ ,
\end{align} 
where $\rho_S(x,x',t)=\<x|\rho_S(t)|x'\>$ is the position representation of the reduced density operator.
The Wigner function of a damped harmonic oscillator is found to satisfy the exact quantum master  equation \cite{hu92}, 
\begin{align}
	\frac{\partial}{\partial t} W(q,p,t)=-\frac{p}{M}\frac{\partial}{\partial q}W+\frac{\partial}{\partial p}[(\gamma p + M \omega^2 q )W] \nonumber \\
		+  D_{pp}(t)\frac{\partial^2}{\partial p^2}W + D_{qp}(t)\frac{\partial^2}{\partial q \partial p}W .\label{Eq:7}
\end{align}
The detailed expressions for the diffusion coefficients $D_{pp}$ and $D_{qp}$ can be found  in Ref.~\cite{hu92}.
The reduced stationary phase space distribution for the Ohmic model is given by the Gaussian Wigner function, 
\begin{align}
	W(q,p) = \frac{1}{2\pi\sqrt{\langle q^2\rangle\langle p^2\rangle}} \exp\left[-\frac{q^2}{2\langle q^2\rangle}-\frac{p^2}{2\langle p^2\rangle}\right]\,,\label{Eq:8}
\end{align}
with exact position and momentum dispersions \cite{gra84},
\begin{align}
	\< q^2\>  &= \frac{\hbar}{M\pi} \sum_{j=1}^3 \left[\frac{(\lambda_j-\omega_D) \, \psi\left(1+\frac{\beta\hbar\lambda_j}{2\pi}\right)}
		{(\lambda_{j+1}-\lambda_j)(\lambda_{j-1}-\lambda_j)}\right]  \notag\\ &+ \frac{1}{M \beta \omega^2} \ ,\label{Eq:9}\\
	\< p^2\>  &=\frac{\hbar M\gamma\omega_D}{\pi} \sum_{j=1}^3 \left[\frac{\lambda_j \, \psi\left(1+\frac{\beta\hbar\lambda_j}{2\pi}\right)}
		{(\lambda_{j+1}-\lambda_j)(\lambda_{j-1}-\lambda_j)}\right] \notag\\
		&+ M^2\omega^2\ew{q^2}  (\eta, M)\label{Eq:10}	\ . 
\end{align}
The parameters $\lambda_i$ are the characteristic frequencies of the damped oscillator and $\psi$ is the digamma function. In the limit of large cutoff frequencies, $\omega_D\gg \gamma, \omega$, the characteristic frequencies can be approximated  by \cite{gra84},
\begin{align}
	\lambda_{1,2} &\simeq \frac{\gamma}{2}\pm \sqrt{ \frac{\gamma^2}{ 4}-\omega^2} \ ,\nonumber\\
	\lambda_3&= \omega_D-\gamma\label{Eq:11}\ .
\end{align} 
A direct consequence of \eq{Eq:10} is that equipartition is in general violated, $M\omega^2 \<q^2\> < \<p^2\>/M$, and  $\rho_S$ is not of the Gibbs form. This can be understood by noting that due to the form of the coupling \eq{Eq:4}, the bath can be seen as continuously measuring the position of the system \cite{wei99}. As a result, the variance of the position  is reduced. The momentum dispersion  then increases following the Heisenberg uncertainty relation. For a vanishingly coupled oscillator, the variances reduce to
\begin{align}
	M\omega^2\< q^2\>_{\gamma=0} =\frac{ \< p^2\>_{\gamma=0}}{M}=\frac{\hbar\omega}{2} \coth\left(\frac{\beta \hbar \omega}{2}\right)\ .
\end{align}
Equipartition therefore holds and the reduced stationary state  is Gibbsian.
The position representation of the reduced density operator is further  given by \cite{wei99},
\begin{align}
	\rho_S(q,q')=\frac{1}{\sqrt{2\pi\<q^2\>}}\exp{\left[-\frac{(q+q')^2}{8\<q^2\>}-\frac{(q-q')^2}{2\hbar^2/\<p^2\>}\right]}\ .\label{Eq:a}
\end{align}
\section{HMF for damped quantum oscillator}\label{sec:4}
The exact expression for the quantum HMF \eqref{Eq:2} for the damped oscillator can be obtained in the following way \cite{gra84}. Since the stationary state \eq{Eq:8} is a Gaussian, it can be regarded as the equilibrium Wigner function of an effective  harmonic oscillator with Hamiltonian,
\begin{align}
	H^*_S =\frac{p^2}{2M^*} + \frac{1}{2}M^*{\omega^*}^2q^2\label{Eq:12} \ ,
\end{align}
with  mass $M^*$ and frequency $\omega^*$.  The latter reads \cite{gre87},
\begin{align}
	W(q,p) &= \frac{\tanh \left({\beta\hbar \omega^* }/{2} \right)}{\pi \hbar} \exp{\left[\frac{2 H_S^*}{ \hbar \omega^*\coth 	
		\left(\beta\hbar\omega^*/2\right)}\right]}\ .\label{Eq:13}
\end{align}
By comparing the two Wigner functions \eq{Eq:8} and \eq{Eq:13}, one can identify  the effective mass,
\begin{align}
	M^* & = \sqrt{\frac{\langle p^2\rangle}{\langle q^2\rangle}} \frac{\beta \hbar}{2} \frac{1}{\arcoth \left( 2 v \right)}\label{Eq:14}  \ ,
\end{align}
and the effective frequency,
\begin{align}
	\omega^* & = \frac{2}{\beta \hbar} \arcoth \left( 2 \nu \right)\label{Eq:15}\ .
\end{align}
Here $v = \sqrt{\langle p^2\rangle \langle q^2\rangle}/\hbar$ is the phase space volume. In order to estimate the effect of the finite coupling on the reduced state of the system, we now determine the deviation  $\Delta H_S$ given by  \eq{Eq:3}. Combining Eqs.~(\ref{Eq:12}), (\ref{Eq:14}) and (\ref{Eq:15}) for the quantum HMF, we arrive at
\begin{align}
	\Delta H_S=A \frac{p^2}{2}+B\frac{q^2}{2}\label{Eq:16} \ ,
\end{align}
where we have introduced the two coefficients
\begin{align}
 	A&=\sqrt{\frac{\<q^2\>}{\<p^2\>}}\frac{2}{\beta\hbar}\arcoth(2v)-\frac1M\ ,\label{Eq:17}\\
	B&=\sqrt{\frac{\<p^2\>}{\<q^2\>}}\frac{2}{\beta\hbar}\arcoth(2v)-M\omega^2\label{Eq:18}\ .
\end{align}
Equation \eqref{Eq:16} gives the exact expression for deviation  $\Delta H_S$ for the damped quantum oscillator. 
The coefficients $A$ and $B$ are divergent for $2v=1$, since  $\arcoth x$ is not defined at $x=1$. This happens for an undamped harmonic oscillator ($\gamma=0$) in its ground state ($T=0$). However, in this case $A=B=0$, as the system relaxes to a Gibbs state.
\begin{figure}
\centering
\subfigure[]{
\includegraphics[width=1\columnwidth]{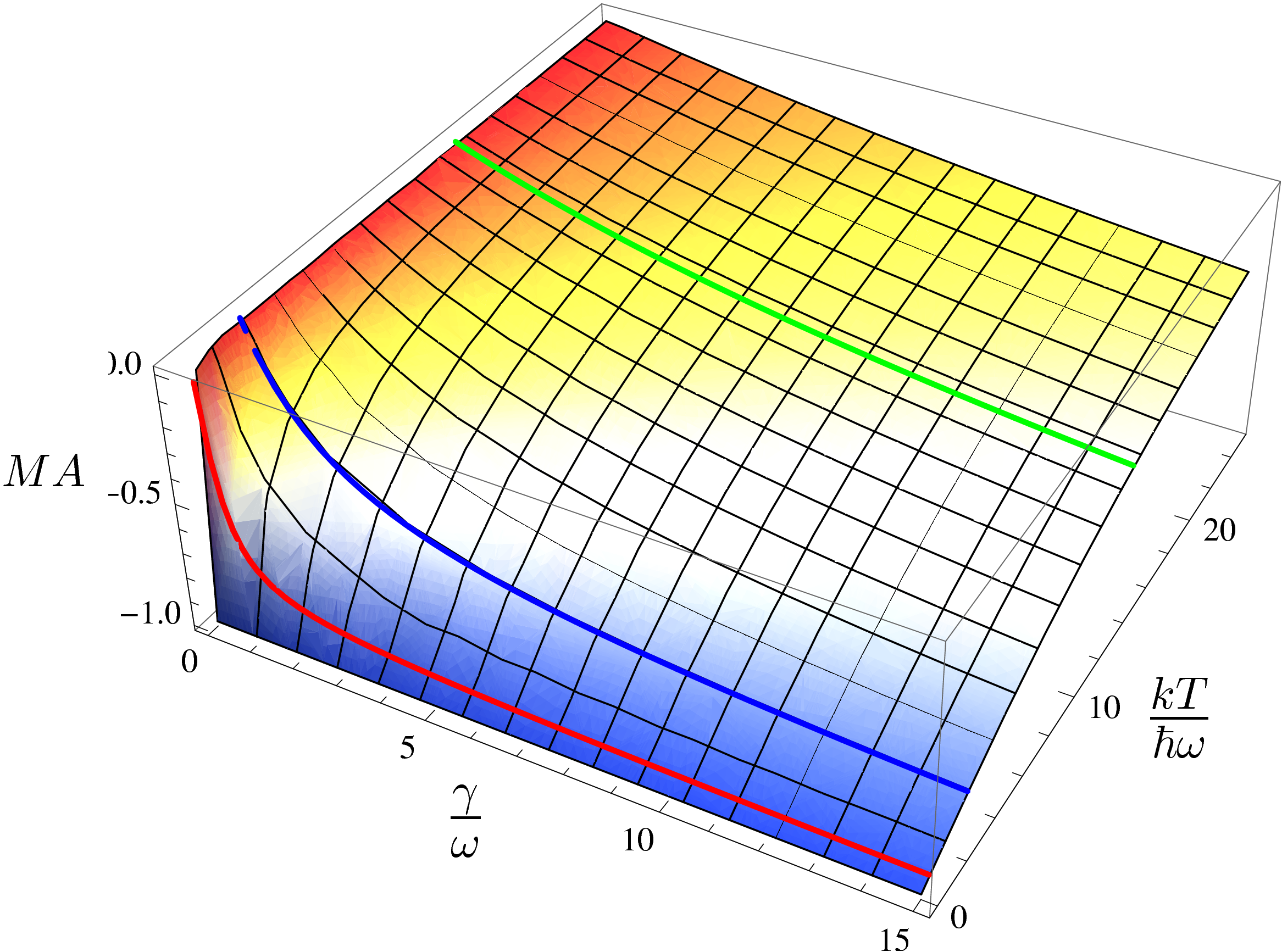}
\label{f1}}
\subfigure[]{
\includegraphics[width=.9\columnwidth]{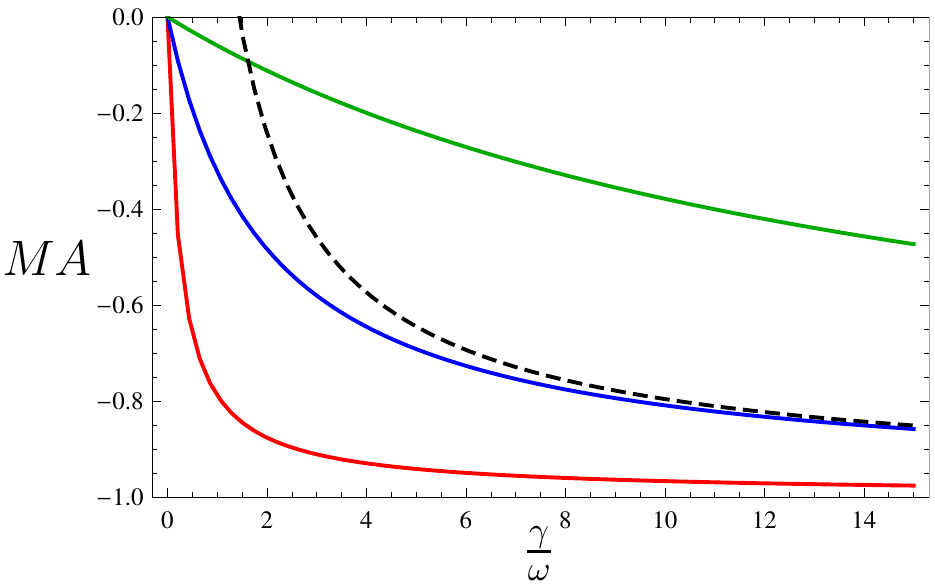}
\label{f1b}
}
\caption{(Color online) (a) Coefficient $MA$, \eq{Eq:17}, as a function of the dimensionless parameters $kT/(\hbar \omega)$ and $\gamma/\omega$. $A$ increases monotonically to zero with temperature and inverse coupling strength. The  deviation from a Gibbs state in momentum is thus maximal in the strong-coupling, low-temperature limit. (b) Cross sections  of the $(T,\gamma)$-curve for $kT/(\hbar\omega)= 0.5$ (red lower line), $ 3$ (blue line), $15$ (green upper line). In the limit of high damping, $A$ coincides with the semiclassical expression given by the quantum Smoluchowski equation (black dashed line for $kT/(\hbar\omega)= 3$), \eq{Eq:33}. Parameters are $\omega_D = 1000$, $\omega = 1$, $M = 1$.}\label{f1}
\end{figure}
\begin{figure}
\centering
\subfigure[]{
\includegraphics[width=1\columnwidth]{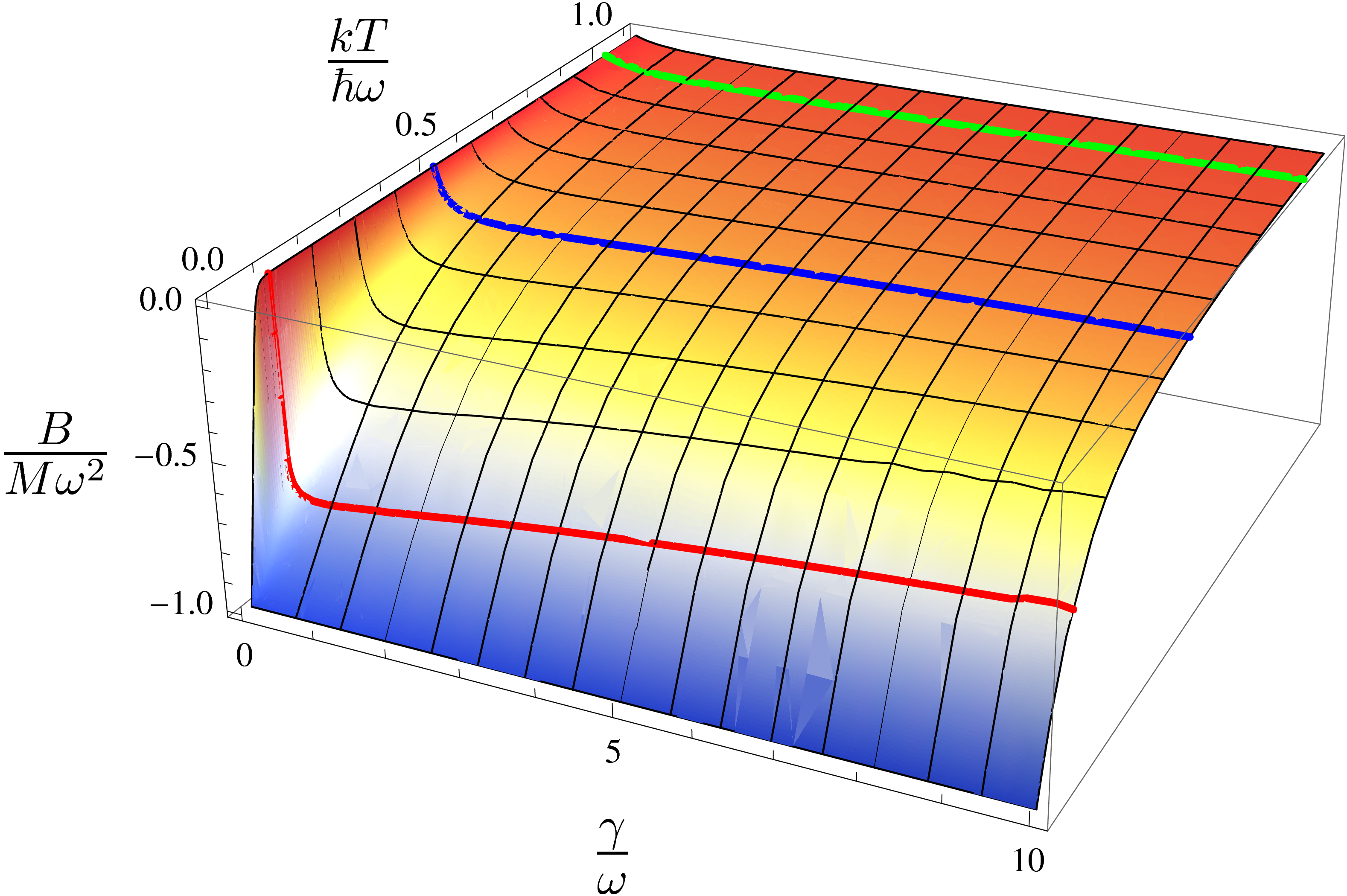}
\label{f2}
}
\subfigure[]{
\includegraphics[width=.9\columnwidth]{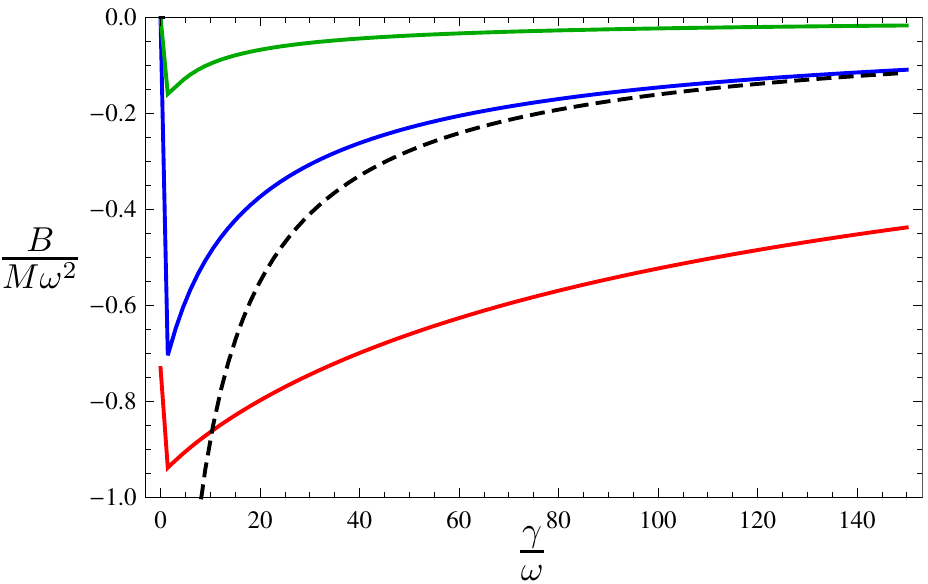}
\label{f3}
}
\caption{(Color online) (a) Coefficient $B/(M\omega^2)$, \eq{Eq:18}, as a function of the dimensionless parameters $kT/(\hbar\omega)$ and $\gamma/\omega$.  $B$ increases monotonically  to zero with temperature. However, it exhibits a minimum as a function of the coupling strength. (b) Cross sections of the $(T,\gamma)$-curve for $kT/(\hbar\omega)= 0.1$ (red lower line), $ 0.5$ (blue line), $1$ (green upper line). For increasing temperatures, the dip gets less pronounced and eventually disappears. In the limit of high-damping, $B$ coincides with the semiclassical expression given by the quantum Smoluchowski equation (black dashed line for $kT/(\hbar\omega)= 0.1$), \eq{Eq:Smolu}. Same parameters as in Fig.~\ref{f1}.}
\end{figure}

\begin{figure}
\centering
\includegraphics[width=.9\columnwidth]{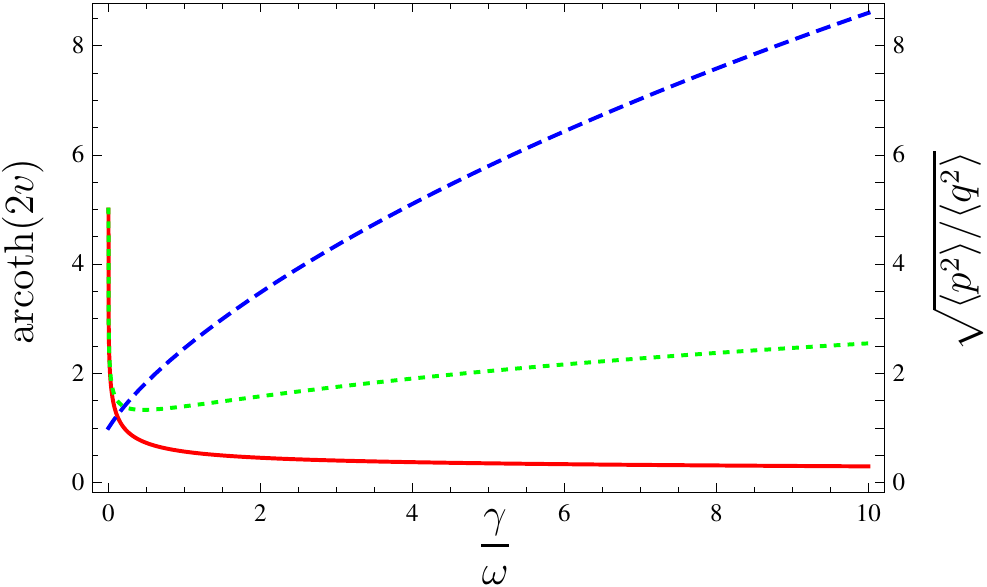}
\caption{(Color online) Function $\arcoth(2v)$ (red continuous) and $\sqrt{\<p^2\>/\<q^2\>}$ (blue dashed) appearing in the definition of the coefficient $B$, \eq{Eq:18}, as a function of the dimensionless parameter $\gamma/\omega$.  Their different behaviors lead to a minimum for their product (green dotted line). Same parameters as in Fig.~\ref{f1} and $kT/\hbar\omega=0.1$.}\label{f3b}
\end{figure}

Figure~\ref{f1} shows the coefficient $A$, \eq{Eq:17}, as a function of  temperature and coupling strength. $A$ is always negative, since $-1\leq M A\leq0$.  We observe that  $A$ decreases monotonically with the coupling and the inverse temperature. The maximum deviation from the Gibbs state is thus achieved in the low-temperature, strong coupling limit. Moreover, for weak coupling, $A$ vanishes rapidly with growing temperature, whereas it decays much more slowly  in the limit of strong coupling, indicating that the deviation from a Gibbs state persists at much higher temperatures in the latter. 
 The coefficient $B$, \eq{Eq:18}, is plotted in Fig.~\ref{f2} as a function of temperature and coupling strength. It is also negative with $-1\leq B/(M\omega^2)\leq0$. While $B$ shows a similar temperature dependence as the coefficient $A$, its dependence on the coupling strength differs significantly. The three  lines highlighted in Fig.~\ref{f2}, and reproduced for clarity in Fig.~\ref{f3},  show that $B$ possesses a minimum. Hence, the deviation of the Gibbs state in position increases up to a certain value of $\gamma$ before decreasing even though the coupling becomes larger. 
The presence of a minimum can be understood by noting that   $\<q^2\>$ decreases with increasing coupling constant, while $\<p^2\>$ increases \cite{wei99}. As a result, the ratio  $\<p^2\>/\<q^2\>$ increases faster than the product  $\<p^2\>\<q^2\>$. The hyperbolic arccotangent of the phase space volume in \eq{Eq:18}, therefore, decreases for growing coupling constants. Its  product with the increasing square root in front thus displays a minimum (see Fig.~\ref{f3b}). Figure \ref{f3} additionally indicates that the minimum gets less pronounced for higher temperatures and eventually disappears in the classical limit.  
Finally, in Fig.~\ref{f4}, we show  the average relative deviation $\<\Delta H_S\>/\<H_S\>$, which vanishes in the high-temperature limit. It is worth noticing that this behavior is peculiar  to the linear coupling  model that we  consider here. For the  case of nonlinear system-bath coupling, the deviation exists even in the classical limit \cite{gel09}. Since $A\<p^2\>$ is in general much larger than $B\<q^2\>$ (see below), the effect we have observed for $B$ is suppressed, and the deviation decreases monotonically with increasing coupling strength. 
In the following, we derive analytic expressions for the coefficients $A$ and $B$ in various limits of interest, including the limits of  low and high temperatures, weak and strong couplings. In all cases, we assume the cutoff frequency $\omega_D$ to be large.

\subsection{High-temperature limit}
\begin{figure}
\centering
\includegraphics[width=1\columnwidth]{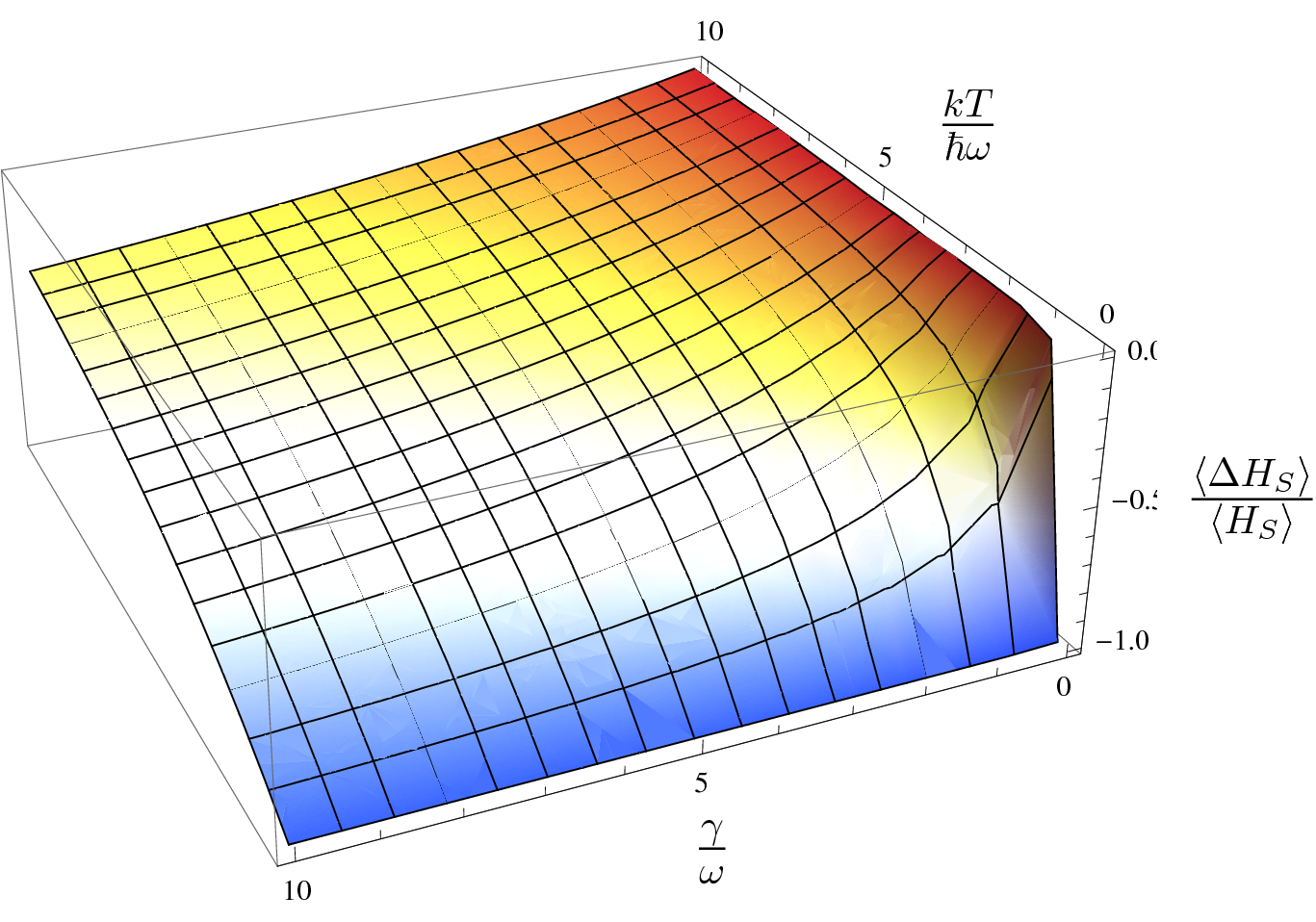}
\caption{(Color online) Relative mean deviation $\<\Delta H_S\>/\<H_S\>$, \eq{Eq:16}, showing the departure from a  Gibbs state as a function of dimensionless temperature and coupling strength. The maximum deviation is observed in the low-temperature, strong-coupling regime. Same parameters as in Fig.~\ref{f1}.}\label{f4}
\end{figure}

In the high-temperature limit, $\hbar\omega \ll kT$, the position and momentum quadratures \eqref{Eq:9} and \eqref{Eq:10} can be expanded to lowest order to yield \cite{gra84},
\begin{align}
	\<q^2\>&\simeq \frac{k T}{M\omega^2}\left[1+\frac{1}{12}\left(\frac{\hbar\omega}{k T}\right)^2\right] \ ,\label{Eq:23}\\
	\<p^2\>& \simeq M k T\left[1+\frac{1}{12}\frac{\hbar^2(\omega^2+\gamma\omega_D)}{(kT)^2}\right]\label{Eq:24}\ ,
\end{align}
where we have made use of the series expansion of the digamma function \eq{Eq:20}. We note that the coupling constant $\gamma$  only appears  in second order in $(\hbar\omega/kT)$. The two variances  Eqs.~(\ref{Eq:23}) and (\ref{Eq:24}), hence, become damping independent in the classical limit, and   equipartition holds, $\<p^2\>/M=M\omega^2\<q^2\>=kT$. For high temperatures, the phase space volume $v$ is large and the $\arcoth$ in the coefficients $A$ and $B$, Eqs.~(\ref{Eq:17}) and (\ref{Eq:18}), can be approximated with the help of \eq{Eq:21}. We obtain,
\begin{align}
 	A&\simeq-\frac{1}{12 M}\frac{\hbar^2(\omega^2+\gamma\omega_D)}{(kT)^2}\label{Eq:25}\ ,\\
	B& \simeq-\frac{M \omega^2}{12}\left(\frac{\hbar\omega}{kT}\right)^2   \label{Eq:26}\ .
\end{align}
Both coefficients  approach zero quickly as $-1/T^2$, see Figs.~\ref{f1} and \ref{f2}. However, unlike $B$, the coefficient $A$ depends explicitly on  $\gamma$ and $\omega_D$. This feature follows from the Ohmic nature of the damping, as seen from \eq{Eq:24} for   $\<p^2\>$. In the limit of high cutoff frequency,  $A$ is much larger than $B$. Combining  Eqs.~(\ref{Eq:23})-(\ref{Eq:26}), we find that  the total deviation $\Delta H_S\sim -1/T^2$ in the classical limit, see Fig.~\ref{f4}. In this regime, the  stationary state of the system reduces to a classical Gibbs distribution, $W(q,p)={\mathcal N}\exp[-(M \omega^2q^2+p^2/M)/(2kT)]$.

\subsection{Low-temperature limit}
For low temperatures, $kT\ll \hbar\omega$, we can approximate the digamma function in  Eqs.~(\ref{Eq:9}) and (\ref{Eq:10}) with the help of \eq{Eq:19}. To proceed further, it is important to distinguish between weak and strong coupling limit. For weak coupling, $kT\ll\hbar\gamma\ll \hbar\omega$, an expansion up to second order in $kT/\hbar\omega$ and first order in $\gamma/\omega$ yields \cite{gra84},
\begin{align}
 	\<q^2\>&\simeq\frac{\hbar\omega}{2M\omega^2}\left[1-\frac{\gamma}{\omega}\left(\frac{1}{ \pi}-\frac{2\pi}{3}\left(\frac{kT}{\hbar\omega}\right)^2\right)\right] \ ,
		\label{Eq:27}\\
	\<p^2\>&\simeq\frac{\hbar\omega M}{2}\left[1-\frac\gamma\omega\left(\frac{1}{\pi}-\frac 2\pi\ln\frac{\omega_D}{\omega}\right)\right]\label{Eq:28}  \ .
\end{align}
For $\gamma=0$, Eqs.~\eqref{Eq:27} and \eqref{Eq:28} lead to the correct ground state   mean energy of  an undamped harmonic oscillator, $\<H_S\>=\<p^2\>/2M+M\omega^2\<q^2\>/2=\hbar\omega/2$.
At the same time, the phase space volume can be approximated  by $v= 1/2(1+C)$, where $C=\gamma(\ln(\omega_D/\omega)-1)/(\omega\pi)$. Therefore, making  use of \eq{Eq:22}, we find the coefficients,
\begin{align}
 	A&\simeq-\frac 1M\left[\frac{k T}{\hbar \omega}(1-Y)\ln (C/2)+1\right]\label{Eq:29} \ ,\\
	B&\simeq-M\omega^2\left[\frac{kT}{\hbar \omega}(1+Y)\ln (C/2)+1\right]\label{Eq:30} \ ,
\end{align}
with $Y=\gamma\ln(\omega_D/\omega)/(\omega\pi)$.
Here, in contrast to the high-temperature limit,  both $A$ and $B$ increase with $T$. As a result, the deviation from the Gibbs state decreases with increasing temperature, see Figs.~\ref{f1} and \ref{f2}. At zero temperature, the coefficients $A$ and $B$ attain their minimum values, $A=-1/M$ and $B=-M\omega^2$, that correspond to the maximum deviation from the Gibbs state.

In the opposite limit of strong damping,  $kT\ll \hbar\omega\ll\hbar\gamma$, a lowest order expansion of the variances, Eqs.~(\ref{Eq:9}) and (\ref{Eq:10}) leads to \cite{gra84},
\begin{align}
	\ew{q^2}&=\frac{2 \hbar}{\pi M\gamma}\ln\frac{\gamma}{\omega}
			+\frac{\pi\hbar\gamma}{3 M \omega^2} \, \left(\frac{kT}{\hbar\omega}\right)^2 \ ,\label{Eq:31}\\
	\ew{p^2}&=\frac{\hbar M \gamma}{\pi}\ln\frac{\omega_D}{\gamma} \ ,\label{Eq:32}
\end{align}
In this regime, the phase space volume $v$ is large and the $\arcoth$ in $A$ and $B$, Eqs.~(\ref{Eq:17}) and (\ref{Eq:18}), can therefore be approximated by \eq{Eq:21}. We then obtain in first order,
\begin{align}
	A&=\frac 1M\left[\frac{k T}{\hbar\omega}\frac{\omega}{\gamma}\frac{\pi}{\ln(\omega_D/\gamma)}-1\right]\label{Eq:33}\\
	B&=M \omega^2\left[\frac\gamma\omega\frac{kT}{\hbar\omega}\frac{\pi}{2\ln(\gamma/\omega)}-1\right]\label{Eq:34} \ .
\end{align}
Again the two coefficients increase linearly with $T$, see Figs. \ref{f1} and \ref{f2}. 
%
%
\section{Semiclassical limit}
In the overdamped  regime, the dynamics of the quantum particle simplifies considerably. In the limit $\gamma/\omega^2\gg (\hbar \beta, 1/\gamma)$, the off-diagonal matrix elements of the reduced density operator of the system in the coordinate representation are strongly suppressed, and a semiclassical description becomes possible. On the coarsed-grained time scale, $t\gg 1/\gamma$, the diffusion coefficients in the master equation \eqref{Eq:7} simplify to \cite{paz09,dil09},
\begin{align}
D_{qp}&=V''(q)\ew{q^2}-\frac{\ew{p^2}}{M}\label{Eq:Dqp}\ ,\\
D_{pp}&=\gamma\ew{p^2}\label{Eq:Dpp}\ .
\end{align}
The position and momentum variances for the harmonic oscillator are given in this limit by \cite{ank01},
\begin{align}
\ew{q^2}&=\frac{1}{\beta M\omega^2}+\lambda\label{Eq:qSmolu}\ ,\\
\ew{p^2}&=\frac{\hbar M\gamma}{\pi}\ln\frac{\omega_D}{\gamma}=\Omega\label{Eq:pSmolu}\ ,
\end{align}
where $\lambda=\hbar\log\left[\hbar\beta\gamma/(2\pi)\right]/(\pi M\gamma)$ measures the strength of quantum fluctuations. The master equation  \eqref{Eq:7} hence  reduces to,
\begin{align}
	\frac{\partial}{\partial t} W(q,t)=-\frac{p}{M}\frac{\partial}{\partial q}W+\frac{\partial}{\partial q}\left[V'(q,t)+\gamma p\right]W\nonumber\\
	+\gamma\ew{p^2} \frac{\partial^2}{\partial p^2}+\frac{\partial^2}{\partial p\partial q}\left[\frac{D(q)}{\beta}-\frac{\ew{p^2}}{M}\right]W\ ,\label{Eq35}
\end{align}
with $D(q)= 1+\beta\lambda V''(q)$. As discussed in Refs. \cite{mac04,luc05}, the effective diffusion
coefficient $D(q)$  should be regarded, for thermodynamic consistency,  as the first order expansion of $D(q)=(1-\beta\lambda V''(q))^{-1}$. Equation \eqref{Eq35} has been derived for arbitrary potentials $V(q)$ in Ref.~\cite{ank03} (note the presence of the incorrect effective potential $V_\text{eff}$ in the  latter \cite{ank01}).  The stationary solution of \eq{Eq35} reads,
\begin{align}
	W(q,p) = N \exp{\left[-\beta V(q)+\frac{\lambda\beta^2}{2}V'(q)^2-\frac{p^2}{2\Omega}\right]}\ .\label{Eq:36}
\end{align}
For the special case of a harmonic potential, $V(q)=M\omega^2q^2/2$, we can immediately identify the coefficient,
\begin{align}
 	B=-\frac{M^2\omega^4}{kT}\lambda\ ,\label{Eq:Smolu}
\end{align}
where we made use of \eq{Eq:16}.  It is important to realize that the semiclassical equation \eqref{Eq35} is valid both in the high-temperature, $kT \gg \hbar \gamma$, and low-temperature, $kT \ll \hbar \gamma$, regimes.
In the high-temperature limit $\lambda= \hbar^2/(12MkT)$ and Eq.~(\ref{Eq:qSmolu}) reduces to Eq.~(\ref{Eq:23}). 

For general potentials $V(q)$, the deviation from  the Gibbs state  takes the form,
\begin{align}
\Delta H_S= A \frac{p^2}{2}+ \frac{\lambda\beta}{2}V'(q)^2 \ ,
\end{align}
where the coefficient A is given by,
\begin{align}
	A&=\frac{1}{\beta\Omega}-\frac1M\label{Eq33} \ ,
\end{align}
and is independent of the system.
In the semiclassical strong-coupling domain, the deviation from the Gibbs state can thus be determined for arbitrary potentials and not only for the harmonic oscillator. We mention that in  the  high-temperature limit,   the semiclassical Hamiltonian of mean force has been examined for a free particle and  a harmonic dumbbell up to  order $\hbar^2$  in Ref.~\cite{gel09}. We note, moreover, that the deviation in position can be directly obtained from the simpler quantum Smoluchowski equation which follows from the 
the semiclassical phase-space master equation \eqref{Eq35} \cite{dil09,ank01,ank03,cof07},
\begin{align}
	\frac{\partial}{\partial t} W(q,t)=\frac{1}{\gamma M}\frac{\partial}{\partial q}\left[V'(q,t)+\frac 1\gamma D(q)\right]W(q,t)\ ,\label{Eq:35}
\end{align}
where we have introduced the  distribution,
\begin{align}
	W(q,t)= \rho_S(q,q,t)=\int d p W(q,p,t)\ ,
\end{align}
The corresponding stationary solution is
\begin{align}
	W(q)= N_q \exp{\left[-\beta V(q)+\frac{\lambda\beta^2}{2}V'(q)^2\right]}\ ,\label{Eq:36a}
\end{align}
\section{Physical interpretation of $\Delta H_S$}\label{sec:5}

A  physical interpretation of the difference $\Delta H_S$ between the quantum HMF and the system Hamiltonian can be given by considering the initial  coupling   between  system and  bath. By treating the latter  process as a thermodynamic transformation, both the change of energy and entropy of the quantum system during the coupling can be determined \cite{hil11}.  We assume that system and bath are initially decoupled, and each in  thermal equilibrium  at temperature $T$. The total density operator is  thus given by the direct product, 
\begin{align}
 	\rho=\frac{\exp{(-\beta H_S)}}{Z_S}\otimes\frac{\exp{(-\beta H_B)}}{Z_B}\ .
\end{align}
If the system-bath interaction  is switched on quasistatically, the entropy of the system changes by $\Delta S$ and an amount $Q$ of heat is exchanged with the heat bath. Both can be evaluated by writing the reduced density operator of the system \eq{Eq:2a} in the form,
\begin{align}
 	\rho_S=\exp{[-\beta (H_S+\Delta H_S-F)]}\ ,
\end{align}
where $F=-(1/\beta) \,\ln Z^*$ is the free energy of the system. The von Neumann entropy of the coupled system, $S=-k\mbox{tr}_S\{\rho_S\ln\rho_S\}$, is then, 
\begin{align}
 	TS= (U-F+\< \Delta H_S\>)\ ,
\end{align}
where $U=\<H_S\> =\mbox{tr}_S\{\rho_S H_S\}$ is the internal energy of the coupled system. According to the first law, the  heat exchanged with the bath during the coupling process is $Q=\Delta U-W$, where the work is  equal to the free energy difference between uncoupled and uncoupled system $W=\Delta F$ \cite{for06}. We, therefore, have,
\begin{align}
	Q= kT \Delta S-\< \Delta H_S\>  \ ,
\end{align}
where we have made use of the fact that, before the coupling ($\gamma=0$), $U_{0}=kT S_{0}-F_{0}$. The above result, valid for any quantum dissipative system, shows that the average difference between the HMF and the bare Hamiltonian of the system is just the difference between $kT \Delta S$ and the heat exchange $Q$ during the coupling process. For vanishingly small coupling, we find $Q=kT\Delta S$, in agreement with standard thermodynamics. %
\section{Conclusion}\label{sec:6}
For a damped quantum system deviations from standard thermodynamics occur in the finite coupling regime. With the help of the quantum HMF, we have derived an exact  expression for the deviation from a thermal  Gibbs state for a damped quantum harmonic oscillator coupled to an Ohmic heat bath. We have obtained useful approximations in the limit of high and low temperature, and of weak and strong coupling. In the semiclassical regime, we have, moreover, used the quantum Smoluchowski equation to derive an approximate formula for the HMF valid for any  quantum system. Finally, we have established a connection between the deviation from a Gibbs state and the thermodynamic change of the system that occur during the initial coupling process. Our findings emphasize the importance of the  HMF in the thermodynamic analysis of strong damped quantum  systems.

This work was supported by  the Emmy Noether Program of the DFG (contract No LU1382/1-1) and the cluster of excellence Nanosystems Initiative Munich (NIM).

\appendix
\section{}
We collect, for convenience, the series expansions of the digamma function used in the evaluation of the position and momentum quadratures, Eqs.~(\ref{Eq:9}) and (\ref{Eq:10}), \cite{abr}, 
\begin{align}
 \psi(x)&=\ln x-\frac{1}{2x}-\frac{1}{12x^2}+\mathcal{O}\left(\frac{1}{x^4}\right) && |x|>1\label{Eq:19}
\end{align}
and
\begin{align}
\psi(1+x)&=-C+\frac{\pi^2}{6}x+\mathcal{O}(x^2)&&|x|<1\label{Eq:20}\ ,
\end{align}
where $C$ is the Euler constant. On the other hand, the series expansions of the hyperbolic cotangent, needed  in the computation of the coefficients $A$ and $B$, Eqs.~(\ref{Eq:17}) and (\ref{Eq:18}), are \cite{abr},
\begin{align}
 \arcoth(y)&=\frac{1}{y}+\mathcal{O}\left(\frac{1}{y^3}\right)&& |y|>1\label{Eq:21}
\end{align}
and
\begin{align}
\arcoth(1+y)&\approx -\frac 12\ln\left(\frac{y}{2}\right)&&|y|<1\ .\label{Eq:22}
\end{align}


\end{document}